\documentstyle[preprint,pra,aps,psfig]{revtex}

\renewcommand{\vec}[1]{\mbox{\boldmath $#1$}}

\tightenlines

\begin{document}
\title{WKB approximation for multi-channel barrier penetrability}
\author{K. Hagino$^{1,2}$ and A.B. Balantekin$^{3}$}
\address{$^1$Yukawa Institute for Theoretical Physics, Kyoto
University, Kyoto 606-8502, Japan }
\address{$^2$Department of Physics, Tohoku University, Sendai
980-8578, Japan}
\address{$^3$Department of Physics, University of Wisconsin, 
Madison, WI 53706}

\maketitle

\begin{abstract}

Using a method of local transmission matrix, we generalize 
the well-known WKB formula for a barrier penetrability to 
multi-channel systems. We compare the WKB penetrability 
with a solution of the coupled-channels equations, and 
show that the WKB formula works well at energies well 
below the lowest adiabatic barrier. We also discuss the 
eigen-channel approach to a multi-channel tunneling, 
which may improve the performance of the WKB formula near 
and above the barrier.  

\end{abstract}


\section{Introduction}

The coupled-channels approach has been a standard method in describing 
atomic, molecular, and nuclear reactions involving internal degrees of
freedom \cite{BT98,DHRS98,HRK99,BBGR93,YK03}. 
In nuclear physics, for instance, the coupled-channels method has been 
successfully applied to heavy-ion fusion reactions at energies around 
the Coulomb barrier in order to discuss the effect of couplings
between the relative motion of the colliding nuclei and inelastic
excitations in the target nucleus \cite{BT98,DHRS98,HRK99}. 
At these energies, the fusion reaction takes place by a quantum
tunneling, and the coupled-channels calculations well account for the 
enhancement of the barrier penetrability due to the channel
couplings. 

A difficulty in the coupled-channels calculations, however, is 
that it is sometimes not so easy to obtain a numerically stable solution 
with a controlled accuracy. 
This is particularly the case 
in the presence of closed channels and/or 
when the coupling strength is strong in the classically forbidden 
region. Several methods have been proposed in order to stabilize the 
numerical solution \cite{BBGR93,J78,L84,RO82}. 

In this paper, instead of directly integrating the coupled-channels 
equations with the stabilization techniques, 
we solve them using the WKB
approximation. To this end, 
we employ the method of local transmission matrix, 
which was originally developed by Brenig and Russ in order to stabilize 
numerical solutions of the coupled-channels equations \cite{BBGR93}. 
We solve the equation for the local transmission matrix under the 
semi-classical assumption, and generalize the well-known WKB formula for 
a barrier penetrability for a single channel to coupled-channel
systems.  
Since the penetrability is expressed in a compact form, the resultant 
WKB formula is entirely free from the problem of numerical
instability. 
Moreover, the WKB method 
can be easily applied to systems with a large number of degrees of
freedom, while obtaining a direct solution of 
the coupled-channels equations can be computationally demanding. 
Also, the WKB method is useful in gaining a physical intuition for
the dynamics of multi-channel tunneling. 

The paper is organized as follows. In Sec. II, we set up 
the coupled-channels 
equations and introduce the local transmission matrix. 
We derive the semi-classical expression for the local transmission 
matrix, and 
obtain the WKB formula for a multi-channel penetrability. We apply 
the WKB formula 
to a three-channel problem, and compare the penetrability with the numerical 
solutions of the coupled-channel equations. 
In Sec. III. we discuss the penetrability at energies near and above the 
barrier height. The WKB formula which we derive works when the
multiple reflection of the classical path under the barrier can be 
neglected, that is, at energies 
well below the barrier. We show that the eigen-channel approach can provide 
good prescriptions at higher energies, where the primitive 
WKB formula breaks down. 
We summarize the paper in Sec. IV. 

\section{Multi-channel WKB formula}

Our aim in this paper is to derive the WKB formula for penetrability
for a one
dimensional potential barrier in the presence of channel couplings. 
We consider the following coupled-channels equations: 
\begin{equation}
-\frac{\hbar^2}{2m}\,\frac{d^2}{dx^2}\,u_{nn_0}(x)+\sum_m
\left[V_{nm}(x)+\epsilon_m\delta_{n,m}-E\right]u_{mn_0}(x)=0. 
\end{equation}
Here, $m$ is the mass of a particle, $\epsilon_n$ is the excitation 
energy for the $n$-th channel, and $E$ is the total energy of the
system. $u_{nn_0}(x)$ is the wave function matrix, where $n$ refers to
the channel while $n_0$ specifies the incident channel. 
Notice that we express the wave functions in a matrix form by 
combining $N$ linearly independent solutions of the coupled-channels 
equations, $N$ being the dimension of the coupled equations. 
For the situation where the particle is incident on the barrier from 
the right hand side, the boundary conditions for 
$u_{nn_0}(x)$ are given by, 
\begin{eqnarray}
u_{nn_0}(x)&\to& T_{nn_0}e^{-ik_nx}~~~~~~~~~~~~~~~~~~~(x\to -\infty), 
\label{boundary1}\\
&\to& \delta_{nn_0}e^{-ik_nx}+R_{nn_0}e^{ik_nx}~~~~(x\to \infty), 
\label{boundary2}
\end{eqnarray}
where $k_n=\sqrt{2m(E-\epsilon_n)/\hbar^2}$ is the wave number for the
$n$-th channel. The inclusive penetrability is then obtained as,
\begin{equation}
P=\sum_n\frac{k_n}{k_{n_0}}\,|T_{nn_0}|^2.
\label{P}
\end{equation}

Let us now introduce the local transmission matrix defined by \cite{BBGR93}
\begin{equation}
\vec{\tau}(x)\equiv
\frac{1}{2}\,\left[i\vec{q}(x)^{-1}\vec{u}'(x)+\vec{u}(x)\right], 
\end{equation}
where $\vec{q}(x)=[2m(E-\vec{W}(x))/\hbar^2]^{1/2}$ with 
$W_{nm}(x)=V_{nm}(x)+\epsilon_n\delta_{n,m}$. 
From Eqs. (\ref{boundary1}) and (\ref{boundary2}), the asymptotic form 
of $\vec{\tau}(x)$ reads, 
\begin{eqnarray}
\tau_{nm}(x)&\to& T_{nm}e^{-ik_nx}~~~~~~~~~~(x\to -\infty), 
\label{boundary3} \\
&\to& \delta_{n,m}e^{-ik_nx}~~~~~~~~~~(x\to \infty). 
\end{eqnarray}
Here, we have used the fact $q_{nm}(x)\to k_n\delta_{n,m}$ as $x\to
\pm\infty$. 
It is easy to show that the local transmission matrix $\vec{\tau}(x)$
obeys the equation\cite{BBGR93}, 
\begin{equation}
\vec{\tau}'(x)=-\frac{1}{2}\vec{q}^{-1}(x)\vec{q}'(x)[1+\vec{\rho}(x)]
\vec{\tau}(x)-i\vec{q}(x)\vec{\tau}(x),
\label{tau}
\end{equation}
where
\begin{equation}
\vec{\rho}(x)\equiv
[i\vec{q}^{-1}(x)\vec{u}'(x)-\vec{u}(x)]
\cdot[i\vec{q}^{-1}(x)\vec{u}'(x)+\vec{u}(x)]^{-1},
\end{equation}
is the local reflection matrix\cite{BBGR93}. 

The WKB approximation may be obtained by neglecting $\vec{\rho}(x)$ 
in Eq. (\ref{tau}) \cite{DR79,R03}, that is, 
\begin{equation}
\vec{\tau}'(x)=
-\frac{1}{2}\vec{q}^{-1}(x)\vec{q}'(x)
\vec{\tau}(x)-i\vec{q}(x)\vec{\tau}(x). 
\label{tauWKB}
\end{equation}
A similar equation has been derived by Van Dijk and Razavy
\cite{DR79,R03}, but by using the 
method of variable reflection amplitude (see also Ref. \cite{BK58}). 
Notice that the asymptotic form of the local reflection matrix
$\vec{\rho}(x)$ is \cite{BBGR93}
\begin{eqnarray}
\rho_{nm}(x)&\to& 0~~~~~~~~~~~~~~~~~~~~~(x\to -\infty), \\
&\to& -R_{nm}e^{2ik_nx}~~~~~~~(x\to \infty). 
\end{eqnarray}
Neglecting $\vec{\rho}(x)$ in Eq. (\ref{tau}) is thus equivalent to
ignoring the reflection, that is reasonable in the semi-classical
limit. This, in fact, corresponds to the lowest order of the Bremmer 
expansion \cite{B49,L51,B82,MH96,G74}, where the WKB formula is
obtained by approximating a smooth potential with a series of sharp
potential steps 
\footnote{The local reflection matrix $\vec{\rho}(x)$ satisfies the
equation $\vec{\rho}'=i(\vec{q}\vec{\rho}+\vec{\rho}\vec{q}) 
-(1-\vec{\rho})\vec{q}^{-1}\vec{q}'(1+\vec{\rho})/2$ \cite{BBGR93}. 
One may solve this equation perturbatively assuming that
$\vec{\rho}(x)$ remains small and finding the correction 
to Eq. (\ref{tauWKB}).}. 

For a single channel problem, Eq. (\ref{tauWKB}) can be easily
integrated to yield 
\begin{equation}
\tau(x)=\sqrt{\frac{q({\infty})}{q(x)}}\,e^{-i\int^xq(x')dx'}. 
\end{equation}
For a coupled-channels problem, however, care must be taken in the
integration, since $\vec{q}(x)$ and $\vec{q}'(x)$ do not commute to
each other in general. 
We attempt to solve Eq. (\ref{tauWKB}) by discretizing the coordinate 
with a mesh spacing of $\Delta x$. Replacing the derivative term by a simple
point difference formula, we obtain 
\begin{eqnarray}
\vec{\tau}(x_{n-1})&\sim&
\vec{\tau}(x_{n})+\Delta x\left[i
\vec{q}(x_n)\vec{\tau}(x_n)
+\frac{1}{2}\vec{q}^{-1}(x_n)\vec{q}'(x_n)
\vec{\tau}(x_n)\right], \\
&\sim& 
\left[1+\frac{\Delta x}{2}\vec{q}^{-1}(x_n)\vec{q}'(x_n)\right]
\cdot 
\left[1+i\vec{q}(x_n)\Delta x\right]\vec{\tau}(x_n), \\
&\sim& 
\left[1-\Delta x\,\vec{q}^{-1}(x_n)\vec{q}'(x_n)\right]^{-1/2}
\cdot 
e^{i\vec{q}(x_n)\Delta x}\vec{\tau}(x_n), 
\label{tauWKB2}
\end{eqnarray}
to the lowest order of $\Delta x$. Using 
$\vec{q}(x_{n-1})\sim
\vec{q}(x_{n})[1-\Delta x\,\vec{q}^{-1}(x_n)\vec{q}'(x_n)]$,  
the first factor in Eq. (\ref{tauWKB2}) is transformed to be 
\begin{equation}
[1-\Delta x\,\vec{q}^{-1}(x_n)\vec{q}'(x_n)]^{-1/2}
\sim\frac{1}{\sqrt{\vec{q}^{-1}(x_n)\vec{q}(x_{n-1})}}\cdot 
\frac{1}{\sqrt{\vec{q}(x_n)}}\cdot \sqrt{\vec{q}(x_n)}. 
\end{equation}
Approximating 
$[\vec{q}^{-1}(x_n)\vec{q}(x_{n-1})]^{-1/2}
[\vec{q}(x_n)]^{-1/2}
\sim [\vec{q}(x_n)
\vec{q}^{-1}(x_n)\vec{q}(x_{n-1})]^{-1/2}
=1/\sqrt{\vec{q}(x_{n-1})}$,
we finally obtain
\begin{equation}
\vec{\tau}(x_{n-1})=
\frac{1}{\sqrt{\vec{q}(x_{n-1})}}
\cdot 
e^{i\vec{q}(x_n)\Delta x}
\sqrt{\vec{q}(x_n)}
\vec{\tau}(x_n). 
\end{equation}
Iterating this equation backward from $x=\infty$, we obtain
\begin{equation}
\vec{\tau}(-\infty)=
\frac{1}{\sqrt{\vec{q}(-\infty)}}
\cdot 
\left(\prod_i e^{i\vec{q}(x_i)\Delta x}\right)
\sqrt{\vec{q}(\infty)}. 
\label{taufin}
\end{equation}
Substituting this expression into Eq. (\ref{P}) together with
Eq. (\ref{boundary3}), the WKB approximation to the multi-channel
penetrability reads
\begin{equation}
P=\sum_n\left|\left\langle n\left|
\prod_i e^{i\vec{q}(x_i)\Delta x}\right| n_0\right\rangle\right|^2. 
\label{PWKB}
\end{equation}
This is the main result in this paper. 
Notice that the factor $k_n/k_{n_0}$ does not appear in the WKB
formula, Eq. (\ref{PWKB}). 
In practice, 
one can evaluate Eq. (\ref{PWKB}) by diagonalizing $\vec{W}(x)$ at 
each $x_i$. This yields 
\begin{equation}
P=\sum_n\left|\left\langle n\left|
\prod_i 
\left[\sum_m\left|m(x_i)\left\rangle 
e^{iq_m(x_i)\Delta x}
\right\langle m(x_i)\right|\right]
\right| n_0\right\rangle\right|^2, 
\label{PWKB2}
\end{equation}
where $|m(x)\rangle$ is the eigen-vector of the matrix $\vec{W}(x)$
with the eigen-value of $\lambda_m(x)$, and 
$q_m(x)\equiv \sqrt{2m(E-\lambda_m(x))/\hbar^2}$. 
For a one dimensional problem, Eq. (\ref{PWKB}) is reduced to the
familiar WKB formula, 
\begin{equation}
P(E)=
\exp\left[-2\int^{x_1}_{x_0}dx'\sqrt{\frac{2m}{\hbar^2}
(V(x')-E)}\right], 
\end{equation}
where $x_0$ and $x_1$ are the inner and the outer turning points,
respectively. 

Let us now apply the WKB formula (\ref{PWKB}) to a three-level
problem. We consider the following coupling potential: 
\begin{equation}
\vec{W}(x)=\left(\matrix{V(x) & F(x) & 0 \cr
F(x)& V(x)+\epsilon & F(x) \cr 
0 & F(x) & V(x)+2\epsilon \cr }
\right), 
\label{cc}
\end{equation}
with 
\begin{eqnarray}
V(x)&=&V_0e^{-x^2/2s^2}, \\
F(x)&=&F_0e^{-x^2/2s_f^2}. 
\end{eqnarray}
The parameters are chosen following Ref. \cite{DLW83} to be 
$V_0$=100 MeV, $F_0$=3 MeV, and $s=s_f=$3 fm, which mimic 
the fusion reaction between two $^{58}$Ni nuclei. 
The excitation energy $\epsilon$ and the mass $m$ are taken to be 
2 MeV and 29$m_N$, respectively, 
where $m_N$=938 MeV is the nucleon mass. 
With these parameters, the three eigen-barriers $\lambda_i(x)$, which 
are obtained by diagonalizing $\vec{W}(x)$ at each $x$, 
have the barrier height of 97.31, 102.0, and 106.7 MeV, respectively, 
while the barrier height for the uncoupled barrier $V(x)$ is 100 MeV 
(See Fig.1). 

Before we study the three channel problem, let us first examine the validity
of the WKB approximation for a single channel case to see whether the
semi-classical approximation works in principle for the parameters
which we choose. Figure 2 shows a comparison between 
the WKB penetrability for the uncoupled
barrier $V(x)$ (the dashed line) obtained with Eq. (\ref{PWKB}) and 
the exact solution. It is plotted in the linear and logarithmic scales 
in the upper and the lower panels, respectively. One clearly sees that the WKB
approximation indeed works well at energies about 2 MeV below the barrier
height and lower. 

As it is well known, the naive WKB approximation breaks down 
around the barrier. 
In fact, the WKB penetrability is unity at the barrier height, 
while the exact result is about a half. 
Around these energies, one needs to improve the WKB formula by using 
the uniform approximation in order to take into account the multiple 
reflection of the classical path under the barrier \cite{BS83,B85}. 
We will discuss this problem more in the next section in connection to the 
penetrability in a coupled-channels system. 

Let us now solve the coupled-channels problem, Eq. (\ref{cc}). 
We integrate Eq. (\ref{PWKB2}) from $x=-15$ fm to $x$=15 fm with 
$\Delta x$=0.05 fm. 
The dashed line in Fig. 3 shows the penetrability in the WKB
approximation for this problem, which is compared to the exact
solution of the coupled-channels equations (the solid line). 
The figure also shows the penetrability for the uncoupled barrier as a
comparison (the dotted line). Remarkably, the WKB formula (\ref{PWKB})
reproduces almost perfectly the exact solution 
at energies well below the lowest adiabatic barrier, i.e., 97.31 MeV, 
as in the single channel problem. 

We notice that 
the WKB penetrability increases much more rapidly than the exact
penetrability at energies corresponding to the height of each
eigenbarrier. This is in close analogue to the single channel problem 
shown in Fig. 2. This behaviour may be expected in the eigen-channel
approach discussed in Refs. \cite{DLW83,HTB97,DL87}. 
We will discuss this point in the next section. 

\section{Penetrability near and above the barrier}

In the previous section, we have shown that the multi-channel WKB
formula works remarkably well at energies well below the lowest
eigenbarrier. Therefore, it can be expected that the WKB formula
provides a useful framework in discussing, for instance, 
the role of inelastic excitations in the colliding nuclei in 
nuclear reactions at extremely low energies, such as astrophysically
relevant reactions, where the standard coupled-channels
calculations may be difficult to carry out. 

At higher energies, however, we found that the agreement between 
the primitive WKB formula and the exact solution of the coupled-channels 
equations becomes poor. For a single-channel problem, 
one can cure this problem by using the uniform approximation \cite{BS83,B85}. 
The WKB formula which is valid at all energies is given by,
\begin{equation}
P(E)=\frac{1}{1+
\exp\left[2\int^{x_1}_{x_0}dx'\sqrt{\frac{2m}{\hbar^2}
(V(x')-E)}\right]},
\end{equation}
where the turning points $x_0$ and $x_1$ become complex numbers 
when the energy $E$ is above the barrier. 
It is not straightforward at all, however, to extend the uniform
approximation to the coupled-channels problem. 
In this section, we instead present two prescriptions to deal with 
the coupled-channels penetrability at energies near and above the
barrier. 

\subsection{Dynamical Norm Method} 

The first prescription is closely related to 
the dynamical norm method developed in Ref. \cite{THA95}. 
It was argued in Ref. \cite{THA95} that the penetrability may be expressed 
as a product of the penetrability in the adiabatic limit and a
multiplicative factor to it which accounts for the non-adiabatic
effect. The latter factor, which was called the dynamical norm factor,
was evaluated through the imaginary time evolution for an intrinsic degree
of freedom with a classical path obtained with the adiabatic
potential. 

We follow here the same idea as in the dynamical norm method,  
and re-express Eq. (\ref{taufin}) as 
\begin{equation}
\vec{\tau}(-\infty)=
\frac{1}{\sqrt{\vec{q}(-\infty)}}
\left(\prod_i e^{iq_0(x_i)\Delta x}e^{i[\vec{q}(x_i)-q_0(x_i)]\Delta x}
\right)
\sqrt{\vec{q}(\infty)},
\end{equation}
where $q_0(x)=\sqrt{2m(E-\lambda_0(x))/\hbar^2}$ is the local wave
number for the lowest eigen-barrier (i.e., the adiabatic barrier),  
$\lambda_0(x)$. The penetrability $P(E)$ is then given by 
\begin{equation}
P=\sum_n\left|\left\langle n\left|
\prod_i e^{i[\vec{q}(x_i)-q_0(x_i)]\Delta x}\right|n_0\right\rangle\right|^2
\cdot
\exp\left[-2\int^{x_1}_{x_0}dx'\sqrt{\frac{2m}{\hbar^2}
(\lambda_0(x')-E)}\right]. 
\label{DNM}
\end{equation}
The second factor on the right hand side (r.h.s.) of this equation 
is nothing more than 
the WKB penetrability for the adiabatic potential $\lambda_0(x)$. 
At this stage, one may replace it by the exact penetrability for 
the adiabatic potential, $P_{\rm ad}(E)$. 
The first factor on the r.h.s. of Eq. (\ref{DNM}) expresses the
non-adiabatic effect, as in 
the dynamical norm factor
introduced in Ref. \cite{THA95}. 
Notice that 
our formula, (\ref{DNM}), 
is in fact an improvement of the dynamical norm method in
Ref. \cite{THA95}, since the classical path is not assumed from 
the beginning in evaluating the dynamical norm factor. 

The result of the modified dynamical norm method is shown in Fig. 4
by the dashed line. It is evident that the agreement between the WKB
approximation and the exact solution improves significantly 
at energies near and slightly above
the adiabatic barrier, although the method provides essentially the
same result as the original WKB formula, (\ref{PWKB}), at higher energies. 

\subsection{Eigen-channel approach}

The second prescription which we discuss is based on the eigen-channel
approximation. 
In this approach, the penetrability is expressed as a weighted sum of
the penetrability for the eigenbarriers \cite{DLW83,HTB97,DL87}, 
that is, 
\begin{equation}
P(E)=\sum_nw_nP_n(E),
\label{eigen}
\end{equation}
where $P_n(E)$ is the penetrability for the eigen-potential
$\lambda_n(x)$. The weight factors $w_n$ are usually estimated by
assuming that the matrix $\vec{W}(x)$ is independent of $x$ through
the interaction range 
\cite{DLW83,DL87}, while the coordinate dependence of $\vec{W}(x)$ is 
properly taken into account in calculating the penetrability, 
$P_n(E)$. One often takes the barrier position of the uncoupled
barrier, $X_b$, in order to estimate the weight factors \cite{DL87}.  
This leads to (see Eq.(\ref{PWKB2})) 
\begin{equation}
w_n=w_n(X_b)=|\langle n(X_b)|n_0\rangle|^2. 
\end{equation}
This procedure is indeed exact when the intrinsic degree of freedom
has a degenerate spectrum \cite{HTB97,HTBB95,NBT86,E81}, since the
matrix $\vec{W}(x)$ can be diagonalized independent of $x$. When the
intrinsic states have a finite excitation energy, however, 
the unitary matrix which diagonalizes $\vec{W}(x)$ explicitly depends 
on $x$, and 
the results 
may depend strongly on the position where the weight factors are
evaluated. Also the weight factors possess some energy
dependence in general. 
In Ref. \cite{HTB97}, we have explicitly shown for a two
channel problem that the optimum weight factors are considerably
different from those estimated at the barrier position, although their
energy dependence appears to be weak. 
A satisfactory procedure to determine the weight factors has not yet
been found so far when the excitation energy is finite. 

In Fig.3, we have shown that the WKB penetrability approaches to a
constant value at the barrier height of each eigen barrier. 
Assuming that the weight factors are independent of energy, one can
exploit this fact to determine a consistent value of the
weight factors in the eigen-channel approach. For instance, at the
barrier height of the lowest eigenbarrier, $E=B_0$, assuming that the
contribution from the higher barriers is negligible, Eq. (\ref{eigen})
suggests 
\begin{equation}
P(B_0)\sim w_0P_0(B_0) \sim w_0, 
\end{equation}
in the primitive WKB approximation (i.e., without the uniform approximation). 
Therefore, if one evaluates Eq. (\ref{PWKB}) at $E=B_0$, it directly provides 
the weight factor for the lowest eigenbarrier. 
One can repeat this procedure $N-1$ times to determine the weight
factors $w_0, w_1, \cdots, w_{N-2}$: suppose that the weight factors for 
the $k$ lowest eigen-barriers have been determined. 
The weight factor for the $(k+1)$-th
eigenbarrier is then estimated as 
\begin{equation}
w_k=P_{\rm WKB}(B_k)-\sum_{i=0}^{k-1} w_i,
\end{equation}
where $B_k$ is the barrier height of the $(k+1)$-th eigenbarrier, and 
$P_{\rm WKB}$ is the penetrability in the WKB approximation,
(\ref{PWKB}). 
Here, we have used the fact $P_i(B_k)=1$ for $i\leq k$ in the
primitive WKB approximation, and assumed $P_i(B_k)=0$ for $i>k$. 
The weight factor for the highest eigenbarrier $\lambda_{N-1}(x)$ is
evaluated as 
\begin{equation}
w_{N-1}=1-\sum_{i=0}^{N-2}w_i,
\end{equation}
in order to ensure the unitarity. 

We apply this prescription to the three channel problem discussed in
the previous section. The weight factors are evaluated to be 0.5914,
0.3543, and 0.0543 for the lowest, the second lowest, and the highest
eigen barriers, respectively. 
The result of the eigen channel approximation with the weight factors
thus estimated is denoted by the filled circles in 
Fig. 4. The result is indistinguishable from the exact solution of the
coupled-channels equations for all the energy region shown in the
figure. 
We thus conclude that the multi-channel WKB formula which we derive in
this paper provides a consistent way to determine the weight factors in
the eigen-channel approach, and it provides a useful and simple 
prescription to compute the penetrability in the presence of channel
couplings at energies from well below to well above the potential
barrier, as long as the weight factors are slowly varying functions of
energy. 

\section{Summary}

We have extended the well known WKB formula for barrier penetrability
to systems with intrinsic degrees of freedom. 
Applying the formula to a three channel problem, we have explicitly
demonstrated that the WKB formula reproduces very nicely the exact
solution of coupled-channels equations at energies well below the
lowest eigenbarrier, i.e., the adiabatic barrier. 
The WKB formula which we derived is applicable even to systems with a large 
number of degrees of freedom, where the standard coupled-channels
calculation cannot be performed due to a computational limitation. 
Our method may therefore provide a useful framework to discuss, 
e.g., a quantum scattering
problem in the presence of coupling to a heat bath \cite{BA03,CC95}. 
The WKB formula is also useful when one discusses the channel coupling 
effect on the tunneling rate 
at deep subbarrier energies, especially in the presence of
closed channels, since the direct integration of the coupled-channels
equations may suffer from a numerical instability. 
Such interesting problems include heavy-ion fusion reactions at
extremely low energies \cite{HRD03}, electron screening effects in
nuclear astrophysical reactions \cite{SRSC01,L93}, and nuclear
structure effects in astrophysical fusion reactions
\cite{HHB03,NCT97}. 

The WKB formula which we derived neglects the effect of multiple
reflection of a classical path under the potential barrier. 
Such primitive formula breaks down at energies near and above the
adiabatic barrier, as is well known. We discussed two prescription 
to cure this
problem. One is the dynamical norm method, where the WKB formula is
re-expressed as a product of the penetrability for the adiabatic
barrier and a multiplicative factor which accounts for the
non-adiabatic effect. By replacing the adiabatic penatrability in 
the WKB approximation by the
exact one, we have shown that this prescription improves the result 
at energies near and slightly above the adiabatic barrier. 
The second prescription is the eigen-channel approximation, where the
penetrability is given as a weighted sum of the eigen penetrability. 
By applying the WKB formula at energies corresponding to the barrier
height of each eigen barrier, we have presented a consistent
procedure to determine the weight factors. We have shown that this 
prescription works well at energies from well below to well above the
barrier, as long as the energy dependence of the weight factors is 
negligible. This method would provide a useful way to simplify the
coupled-channels calculations in realistic systems. 

For a single channel problem, the primitive WKB formula can be improved
by using the uniform approximation \cite{BS83,B85}, where the
resultant WKB formula is applicable at energies 
both below and above the barrier. 
It will be an interesting, but very challenging, problem 
to extend it to a multi-channel problem. For this purpose, it will be
extremely helpful to construct a solvable coupled-channels model. A
work towards this direction is now in progress \cite{HB04}. 

\section*{Acknowledgments}

The authors thank H. Horiuchi and Y. Fujiwara for useful discussions. 
A.B.B. gratefully acknowledges the 21st Century for Center of Excellence 
Program ``Center for Diversity and Universality'' at Kyoto University for
financial support and thanks the Yukawa Institute for Theoretical
Physics for its hospitality. 
This work was supported in part by the U.S. National Science
Foundation Grant No. PHY-0244384 and in part by
the University of Wisconsin Research Committee with funds granted by
the Wisconsin Alumni Research Foundation.

\begin{figure}
  \begin{center}
    \leavevmode
    \parbox{0.9\textwidth}
           {\psfig{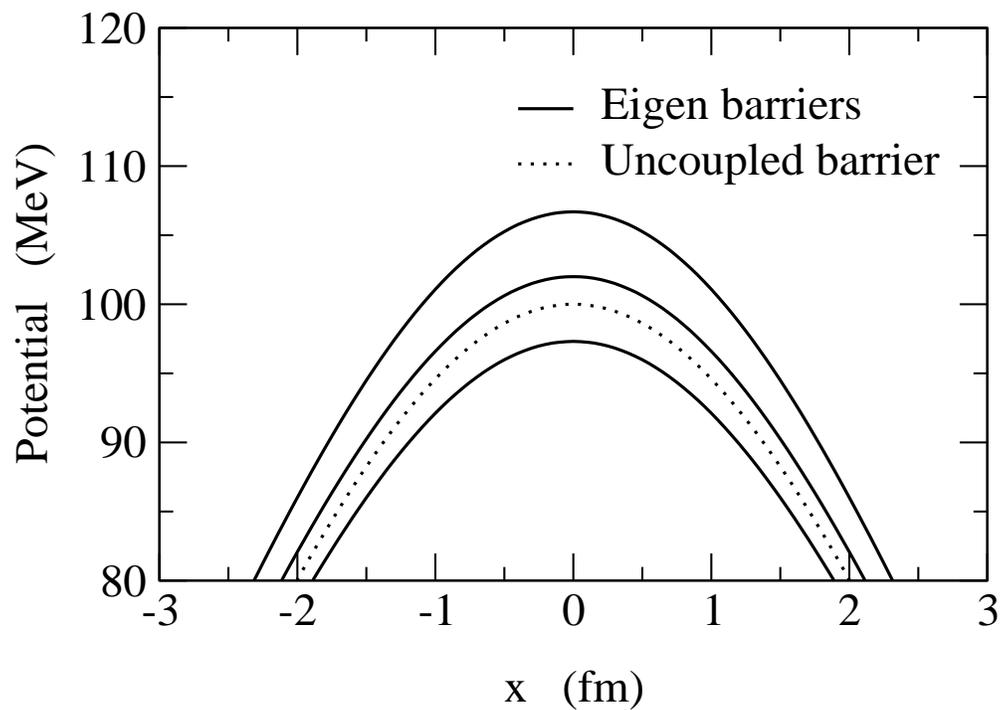}}
  \end{center}
\protect\caption{
The eigenpotentials for the three channel problem obtained 
by diagonalizing the coupling matrix $\vec{W}(x)$ at each $x$ (the
solid lines). The uncoupled barrier $V(x)$ is shown by the dotted line
for a comparison. 
}
\end{figure}

\begin{figure}
  \begin{center}
    \leavevmode
    \parbox{0.9\textwidth}
           {\psfig{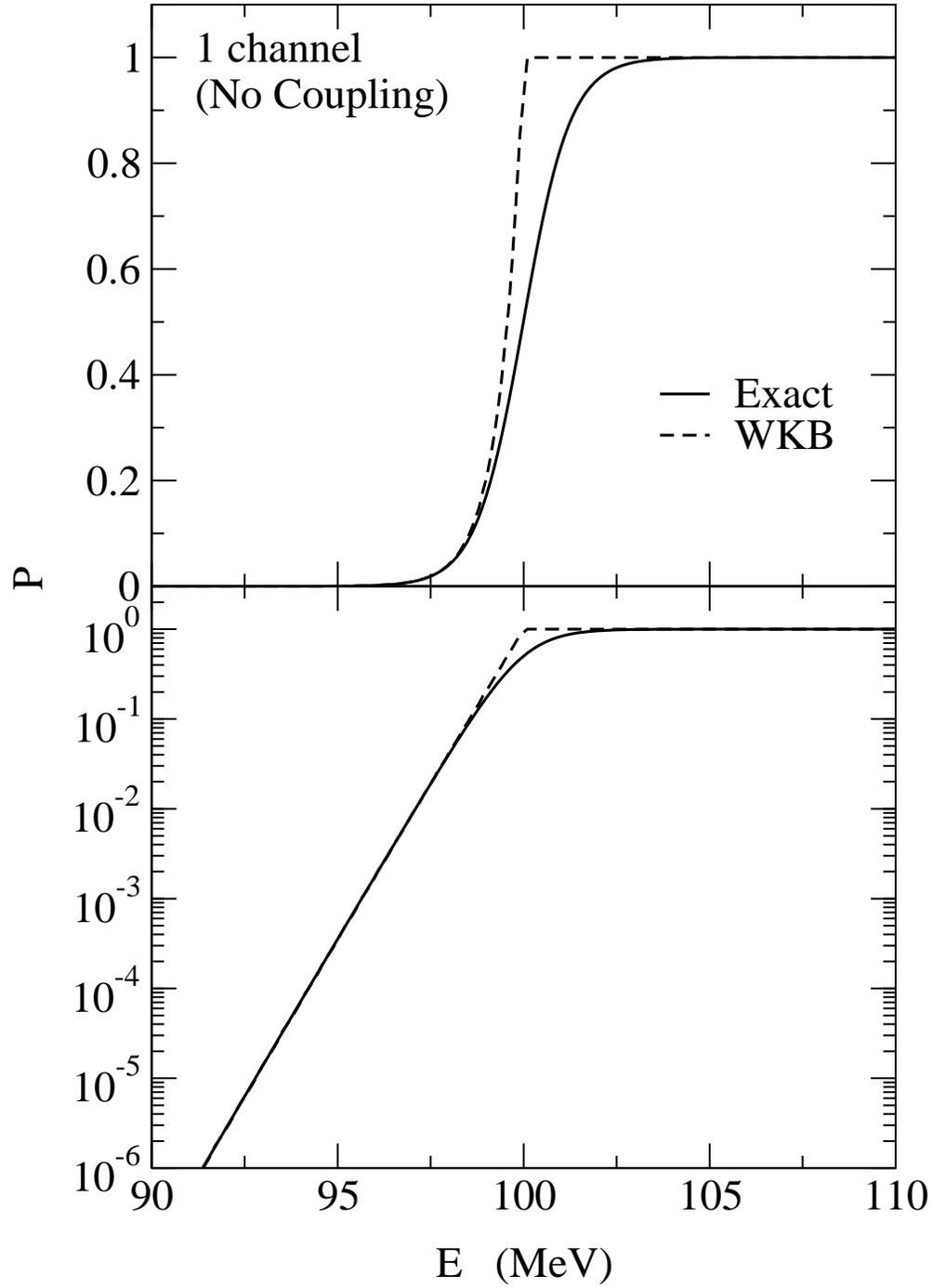}}
  \end{center}
\protect\caption{
The barrier penetrability for the uncoupled barrier $V(x)$ as a
function of energy $E$ in the linear (the upper panel) and the 
logarithmic (the lower panel) scales. 
The solid and the dashed lines are the exact 
solution and the WKB approximation, respectively. 
}
\end{figure}

\begin{figure}
  \begin{center}
    \leavevmode
    \parbox{0.9\textwidth}
           {\psfig{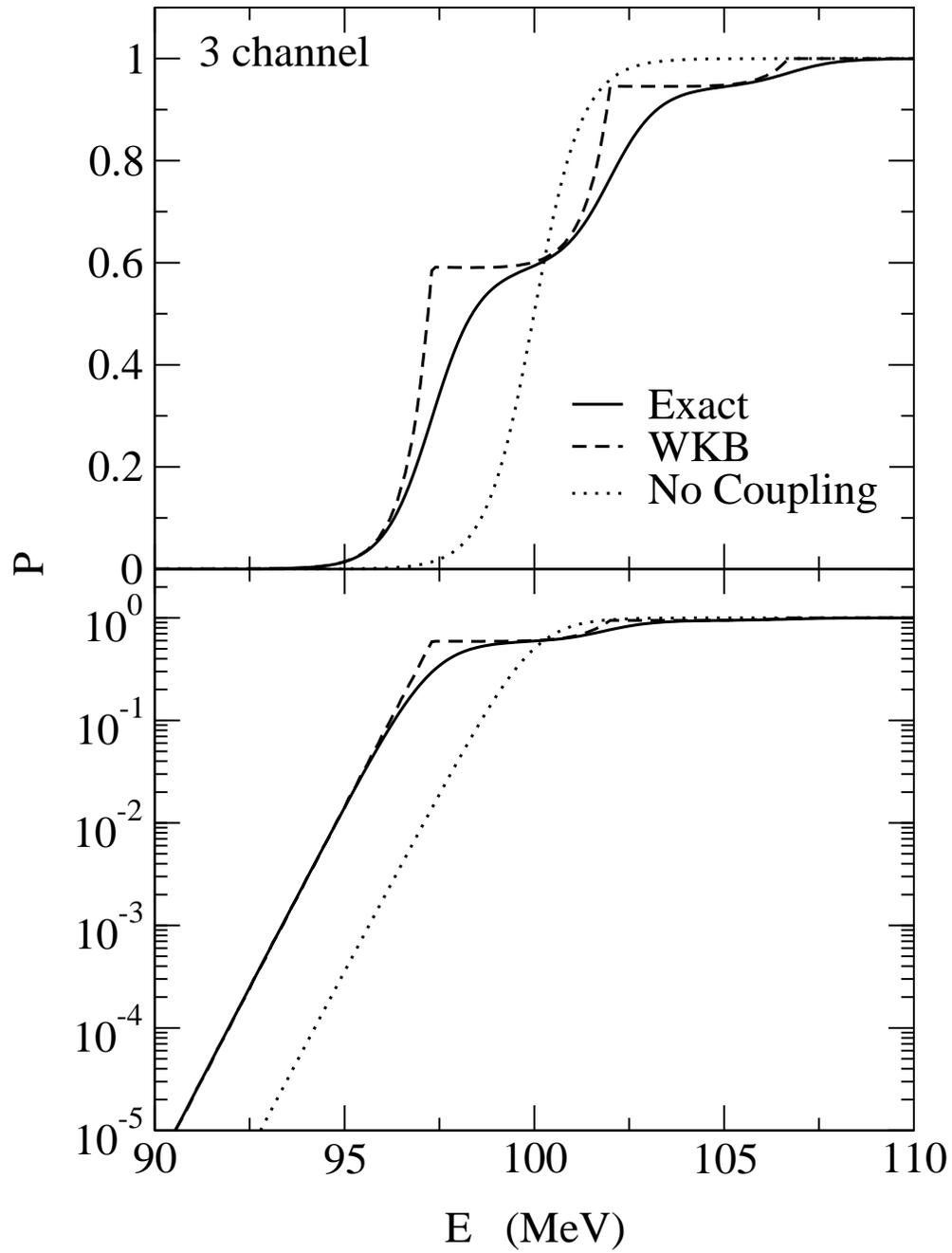}}
  \end{center}
\protect\caption{
Same as fig. 2, but for the three-channel problem. 
The dotted line shows the penetrability in the no coupling limit. 
}
\end{figure}

\begin{figure}
  \begin{center}
    \leavevmode
    \parbox{0.9\textwidth}
           {\psfig{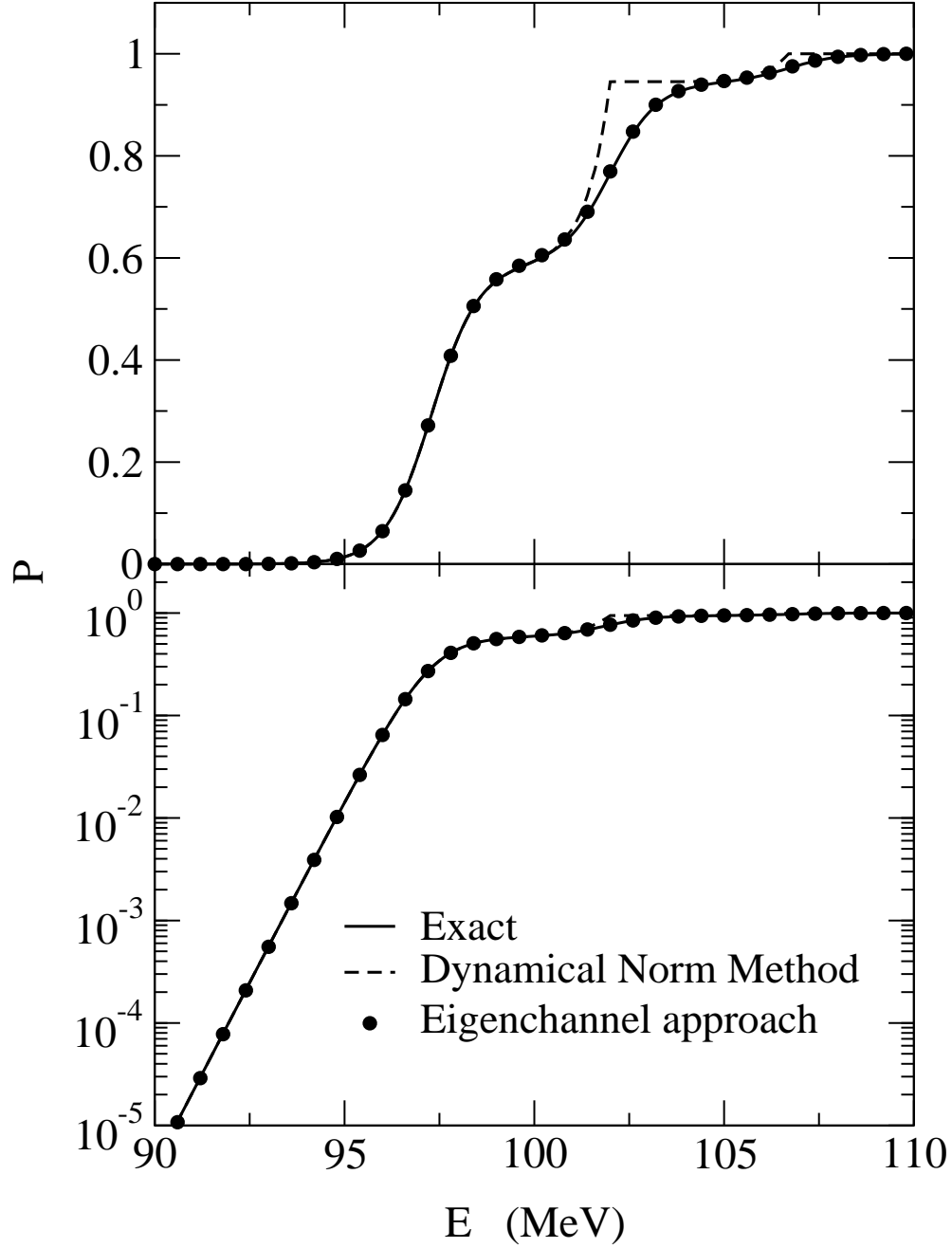}}
  \end{center}
\protect\caption{
Comparison of the barrier penetrability calculated by several methods. 
The solid line shows the exact solution of the coupled-channels
equations. The dashed line is obtained with the modified adiabatic
method which takes the non-adiabatic correction into account, while
the filled circles are results of the eigen-channel approximation. 
The upper and the lower panels shows the penetrability in the linear
and the logarithmic scales, respectively. 
}
\end{figure}

\end{document}